\documentstyle[preprint,aps]{revtex}
\tightenlines 
\begin{document}

\gdef\journal #1, #2, #3, 1#4#5#6{
      {\sl #1~}{\bf #2}, #3 (1#4#5#6)}
\def\SSP{\journal Sol. Stat. Phys., }
\def\PRB{\journal Phys. Rev. B, }
\def\JPC{J. Phys. C, }
\def\PL{\journal Phys. Lett., }
\def\PRD{\journal Phys. Rev. D, }
\def\JPCS{\journal Phys. Chem. Sol., }
\def\NPB{\journal Nucl. Phys. B., }
\def\PR{\journal Phys. Rev., }
\def\PRL{\journal Phys. Rev. Lett., }
\def\JPAMG{\journal J. Phys. A: Math. Gen., }
\def\Tr{\rm Tr}
\def\SSC{\journal Surf. Science, }
\def\ASSP{\journal Adv. in Solid State Phys., }
\def\FNT{\journal Fiz. Nitk. Temp., }
\def\SJLTP{\journal Sov. J. Low Temp. Phys., }
\def\DAN{\journal Dokl. Akad. Nauk., }
\def\Euphys{\journal Europhys. Lett., }
\def\ltwid{\mathrel{\raise.3ex\hbox{$<$\kern-.75em\lower1ex\hbox{$\sim$}}}}
\def\bfx{{\bf x}}
\def\bfy{{\bf y}}
\def\bfp{{\bf p}}
\def\bfq{{\bf q}}
\def\bfk{{\bf k}}
\def\bfK{{\bf K}}
\def\bfr{{\bf r}}
\def\sh{\rm sh}

\title{Periodic Density Modulation Effects on a Correlated two-dimensional Composite
Fermion System}

\author{S. Sakhi}
\address{College of Arts and sciences, American University of Sharjah,\\
Sharjah, UAE, P.O.Box 26666}
\maketitle
\begin{abstract}
We examine theoretically the effects caused by a periodic external potential on
the correlated motion of a two-dimensional electron system under strong magnetic fields corresponding to a filling factor $\frac{1}{2}$. To describe the resulting complex dynamics, we adopt a composite fermion approach and we
determine in a two loop approximation the density-response function $K_{00}(\bfq,\omega)$
and the compressibility. We show explicitly that the long-wavelength limit of $K_{00}(\bfq,\omega)$ exhibits substantial anisotropic behavior induced by the modulation, and that the system tends to be incompressible in a direction orthogonal to the modulation as opposed to its response along the modulation.

\end{abstract}


\footnote{email: ssakhi@aus.ac.ae}
\eject

Many  collective phenomena in two-dimensional electron systems
(2DES)originate from the combined effect of 
strong correlations, high magnetic fields and low temperatures. The fractional quantum Hall                    
effect (FQHE) \cite{PraGir}is one such remarkable state of matter with fractional charges, spins and statistics as well as unprecedented order parameters. At filling factor
$\nu=1/2$ more intriguing features arise, 
including a seemingly metallic dc magneto transport and a surface
acoustic wave anomaly \cite{will1}. 

Over the years the wisdom that has been gained from the study of this intricate planar many-particle behavior is that casting electron-electron correlation in terms of flux attachment facilitates the comprehension of this phenomenon. In this approach an impinging magnetic field is viewed as producing vortices in the 2DES, one for each flux quantum $\phi_0=h/e$. For strong magnetic fields corresponding to fractional filling of the first Landau level, there are more vortices than there are electrons. Placing more vortices onto each electron reduces considerably the Coulomb interaction, and the system becomes more tractable with a viable ground state. 

Mathematically the idea of flux attachment is efficiently embodied by Chern-Simons gauge interactions. In one approach the FQHE, as seen in electron systems, is interpreted as the integer quantum Hall effect (IQHE) experienced by some quasi-particles, `composite fermions,' which are the result of attaching
an even number of flux quanta to each electron. In the mean field
approximation, these composite fermions experience
an effective magnetic field composed of the external field and the
statistical magnetic part. This formulation has been analyzed in
\cite{LopFra}
where the response functions and collective modes were obtained. For a filling factor
$\nu=1/2$, the average effective magnetic field vanishes, and composite
fermions appear to form a fermi surface with coupling to fluctuating gauge
fields.  Within a random phase approximation (RPA) scheme, this case was
analyzed in \cite{Halp}, showing the appearance of
singularities of non-Fermi liquid nature in the electronic self-energy.
Further, the large enhancement in
the effective mass motivated more studies \cite{other}
on this $\nu=1/2$
compressible state. In another approach, an odd number of flux quanta is attached to each electron resulting in composite bosons. These composites reside in apparently zero magnetic field, and they Bose condense into a new ground state with
an energy gap, characteristic for such Bose condensation.This energy gap guaranties the quantization of the Hall resistance and leads to a vanishing resistance. This formulation has been adopted in \cite{LeeZhang} where a Chern-Simons-Landau-Ginzburg theory was analyzed.

In this paper, we explore theoretically the
effects caused by a charge density modulation (CDM) on a $\nu=1/2$ compressible
2DES. Recently, experiments have been devised to implement these periodic density modulations on GaAs/AlGaAs heterostructures \cite{smet}. These studies show the influence of the periodic variation of the density on the dc-transport properties and the propagation of surface acoustic waves (SAW) across the modulated two-dimensional system. 
The SAW experiments showed features at $\nu=1/2$ that resemble those observed at fractional quantum Hall states \cite{will2}. More specifically, in a direction orthogonal to the CDM, the
measured ultrasound velocity shift displays unexpectedly a peak in its
magnetic
field spectrum. The effect is similar to the response of $\nu=1$ and $\nu=1/3$
 Hall
states, suggesting that a charge density modulation applied to a $\nu=1/2$ system
 produces a transition
from a composite fermion filled fermi sea to an anisotropic state, where
the fermi sea properties are preserved in one direction, but the system
acts like a quantum Hall state in the orthogonal direction.

A fundamental property inherent to the flux attachment in the composite fermion picture is that a modulating charge density leads necessarily to a modulating fictitious magnetic field. Consequently, in regions with a surplus of charge carriers the statistical magnetic field overcompensates the external magnetic field (there is a superfluous fictitious flux quanta); whereas in regions with a deficit in carriers there is under compensation between the statistical magnetic field and the external field (missing fictitious flux quanta). Pictorially, along the modulation, composite fermions see as many regions with positive fluxes  as many with negative fluxes; therefore, one would expect in this case a fermi surface behavior.
In the orthogonal direction, however, composite fermions see always fluxes pointing in the same direction. To describe this complex dynamics, we determine, using a Lagrangian approach, the density-response function $K_{00}(\bfq,\omega)$,
and the compressibility in the long-wavelength limit.

In our theoretical investigation we firstly address the effect of weak periodic potentials on those
composite fermion states that do not display near-degeneracy. The energy
shift in the states, as expected by simple perturbations, is quadratic in
the
imposed periodic potential. Taking account of self-energy
and vertex corrections, we then obtain the compressibility, defined
from the density-density
correlation function. The long-wavelength limit analysis of this quantity
reveals quite distinct behaviors, depending on the 
direction of $\bfq$ with respect to the modulation wave vector ($\bfK$).

Next, we address the case of near degenerate states, which are mostly affected by
weak  periodic fields since the energy shift is linear in the field.
We then solve a Dyson-Schwinger-like
equation, and obtain corrections to polarization tensors. The derived
compressibility
displays similar anisotropic behavior as in the first case; however, in the transverse 
direction, the effect is  enhanced due to process involving 
nearly degenerate composite fermions states in which the net momentum changes by the modulation wave vector $\bfK$.

The action describing the 2DES reformulated in terms of composite fermions
is given by

\begin{eqnarray}
\label{action}
S&=&\int_0^{1/T}d\tau \int d\bfx
\psi^\dagger \left(\partial_\tau +ia_\tau-\mu-V(\bfx)
\right)\psi -\frac{1}{2m^*}|{\bf
D}\psi|^2\nonumber\\
&-&\frac{i{\eta}}{2}\int_0^{1/T}d\tau \int d\bfx
\epsilon_{\mu\nu\lambda}a_\mu\partial_\nu
a_\lambda\nonumber\\
&+&\frac{1} {2}\int_0^{1/T} \int d\bfx d\bfy\,
\,\left(|\psi({\bf x},\tau)|^2-\bar{\rho}\right)
v(|{\bf x-y}|)
\left(|\psi({\bf y},\tau)|^2-\bar{\rho}\right),
\end{eqnarray}
here $\mu$ is the chemical potential; $m^*$ is the effective masse.
$D$ is the usual covariant derivative which couples fermions to the sum
of the external gauge field ${\bf A}_{\rm ext}$ and the Chern-Simons
field ${\bf a}$. Only the latter is dynamical with a Chern-Simons
kinetic term; the coefficient $\eta=1/(4\pi)$ is chosen such that two
flux quanta are attached to each electron allowing, in the mean field,
a complete cancellation of the effective magnetic field. 
$V(\bfx)$ is the modulating external potential assumed to be weak.
Finally, the last
term in the action describes the Coulomb repulsion between the charged fermions.

The underlying idea of flux attachment is clearly seen from the
constraint equation obtained by varying action (1) with respect to 
$a_\tau$,

\begin{equation}
|\psi|^2=\eta b,
\end{equation}
where $b$ is the statistical magnetic field. In the mean field, the
effective magnetic field $B_{\rm eff}=B-b/e$ is zero; consequently, 
composite fermions form a fermi sea with residual coupling to gauge
fluctuations $\delta a_\mu$.

To determine the response function 
$K_{\mu\nu}$, we probe the system with an electromagnetic field 
${\tilde A}_\mu$,
and integrate out both fermions and gauge fluctuations $\delta
a_\mu$,
resulting in an effective action functional:

\begin{eqnarray}
&&\exp(-W[{\tilde A}_\mu^\alpha])=\int {\cal D}\psi {\cal D}\psi^\dagger
{\cal D}\delta a_\mu e^{S} \nonumber\\
&=&\int{\cal D}\delta a_\mu \exp\left({{\rm
Tr}\ln{\bigl(\partial_\tau-\epsilon\bigl(-i\nabla-e{\tilde A}-\delta a \bigr)
-V(\bfx)+i\delta a_\tau\bigr)}-\frac{\eta^2}{2}\int \delta b v \delta
b-i\eta\int \delta a_\tau\delta b}\right).
\end{eqnarray}
In obtaining this expression firstly, we used the constraint to simplify
the quartic Coulomb interaction; secondly, we carried out the remaining fermionic integration,
resulting in the familiar fermionic determinant which is then expressed as 
$e^{{\rm Tr}\ln()}$. To carry out the gauge field integration, however, 
we
chose the coulomb gauge (natural gauge for this problem), allowing  us 
to write $\delta
a_i(\bfq,\tau)=i\epsilon^{ij}(q_j/q)a(\bfq,\tau)$; thus, 
the gauge propagator $D_{ab}$ is a $(2,2)$ matrix with
indicies ($a,b=0,1$) denoting respectively the longitudinal and transverse
components. To obtain the
density-density
response function, however, one only needs the transverse component of the
gauge 
field since
\begin{equation}
K_{00}(x,\tau;y,\tau')=\frac {1} {\beta}\sum_{\omega} e^{i\omega
(\tau-\tau')}\int 
\frac{d^2q} {(2\pi)^2} e^{iq(x-y)} (-\eta^2 q^2) D_{11}(q,\omega).
\end{equation}

To second order in the weak external periodic field, the transverse gauge propagator $D_{11}$ 
that includes self-energy and vertex corrections is given by

\begin{equation}
D_{11}=\frac {D_{11}^{(0)}+\Pi_{00}^{(2)}det(D^{(0)})} {1-tr(D\Pi^{(2)})}.
\end{equation}
$D^{(0)}$ is the matrix gauge propagator in the absence of any external
potential. 
In a one-loop approximation and long wavelength limit, its components are
\begin{eqnarray}
D_{00}^{(0)}(q)&=&2\pi/m  \\
D_{01}^{(0)}(q)&=&-\frac {iq} {2m} D_{11}(q)  \\
D_{11}^{(0)}(q)&=& \frac {1} {i\gamma\omega/q+q^2\chi_0(q)}, 
\end{eqnarray}
where $\chi_0(q)=1/(12\pi m)+(v(q)+2\pi/m)/(16\pi^2)$, and $\Pi^{(2)}$, a
$(2,2)$ matrix, accounts of buble
diagrams, including  self-energy and vertex corrections as seen in figure.
In the 
long wavelength limit, however, one only needs the transverse 
component $\Pi^{(2)}_{11}$. The relevant diagrams are depicted in figure
(1); evaluating these terms, and replacing them in (4) and (5), we arrive
at our first result 

\begin{equation}
K_{00}(q)= \frac {1} {\chi_0(q)+C |U_K|^2\sin^2(\theta)/q^2},
\end{equation}
where $\theta$ is the angle formed between the directions of $q$ and $K$; 
$U_K=(2\pi/m)V_K/\chi_0(K)$, the screened external potential and $C$ is a constant that depends on $K$.
The density response function shows clearly anisotropy effects brought about by the modulation. These effects become more pronounced when the interaction between electrons is short range. In fact along the modulation, and in the long wavelength limit $K_{00}(q_{\parallel}, q_{\perp}=0)=1/\chi_0$ indicative of a compressible behavior. Where as in the orthogonal direction $K_{00}(q_{\parallel}=0, q_{\perp})=q_{\perp}^2/C|U_K|^2$ indicative of an incompressible state. As a comparison we give here the long wavelength limit of the density response function for a state at filling factor $\nu=1/3$ that was derived in a Chern-Simons-Landau-Ginzburg-like description \cite{LeeZhang},$K_{00}(q)=\kappa q^2/[q^2+4\pi^2\kappa\rho\/(m\nu^2)]$. We have shown that to second order in the periodic field the system is more incompressible in a direction orthogonal to the modulation as opposed to its response along the modulation.

Next, we consider the contribution of near degenerate states. One needs,
in this case, to calculate the contribution to the polarizations
$\Pi_{11}$ and $\Pi_{00}$ with  internal fermionic propagators
accounting
for degeneracy between states with momentum $\bfp$ and $\bfp+\bfK$.
Solving
a Schwinger-Dyson-like equation in this case, one obtains:
 
\begin{equation} 
G(p-K,p;\omega)=\frac {(i\omega-\zeta(p-K))-U_K(p)}
{[i\omega-E^+(p,K)][i\omega-E^-(p,K)]} 
\end{equation} 
where $\zeta(p)=p^2/(2m)-\epsilon_F$,
$U_K(p_\perp)=U_K\sqrt{1+4p_\perp^2/K^2}$, and $E^{\pm}$
describes the dispersion relation of the two branches near a Brag plane.
They are given by 
\begin{equation}
E^{\pm}=\epsilon(p_\perp)+\frac{\epsilon(p_\parallel)+\epsilon(p_\parallel')}{2}
\pm\sqrt{\left(\frac{\epsilon(p_\parallel)-\epsilon(p_\parallel')}{2}\right)^2+
|U_K(p_\perp)|^2}, 
\end{equation}
with $p_\parallel'=p_\parallel-K$.
The contribution of these states to the polarization $\Pi_{11}$ is
\begin{equation}
\Pi_{11}=\frac{1}{m^2}\int\frac{d^2p}{(2\pi)^2}\frac{f(E^+(p,K))-f(E^-(p,K))}
{E^+(p,K)-E^-(p,K)+i\omega} |\hat{q}\wedge <\psi^{(1)}|e^{-i\bfq.\bfr}
(i\nabla)|\psi^{(2)}>|^2. 
\end{equation}
The matrix element in this
equation, in the long wavelength limit, is $<\psi^{(1)}|e^{-i\bfq.
\bfr}(i\nabla)|\psi^{(2)}>=K/2$. When the fermi level is in a gap between
the
two branches $E^{\pm}(p)$, the polarization evaluates to 

\begin{equation}
\Pi_{11}=-\frac{8\epsilon(K/2)}{\pi}\sin^2(\theta)\frac{|U_K|^2}
{4|U_K|^2+\omega^2} 
\end{equation}

Using this result in the response function, we obtain:  

\begin{equation} 
K_{00}(q,\omega)=
\frac{\eta^2 q^2}{q^2\chi_0(q)+\frac{i\gamma\omega}{q}+\frac{K^2}{e^2
m\pi}\sin^2(\theta)\frac{|U_K|^2}{\omega^2+4|U_K|^2}} 
\end{equation}
Similar to our previous result derived in equation (9) we note the behavior predicted by this formula is rather anisotropic especially for short range interactions between electrons. 
 
In conclusion, we have shown that imposing a modulating charge density on the correlated motion of a two-dimensional electron system at Landau-level filling $\nu=1/2$ changes drastically the system's response to probing electromagnetic fields. Substantial anisotropic behavior is shown explicitly in the calculated density response function, indicating the system is incompressible in a direction orthogonal to the modulation as opposed to its response along the modulation.

\vskip 2cm
\begin{figure}[t]
\caption{Corrections to the gauge field propagator including self-energy and vertex corrections. The wiggly lines represent the gauge field propagator; the cross on the wiggly lines represent the imposed external modulating potential; the solid lines represent the composite fermion propagator}
\end{figure}

\end{document}